\documentclass[aps,pra,reprint,groupedaddress]{revtex4-1}
\usepackage{booktabs}
\usepackage{graphicx}
\usepackage{dcolumn}% Align table columns on decimal point
\usepackage{bm}% bold math
\usepackage{amsmath}
\usepackage{txfonts}
\usepackage{color}
\begin{document}

\newcommand{\tabincell}[2]{\begin{tabular}{@{}#1@{}}#2\end{tabular}}

% Use the \preprint command to place your local institutional report
% number in the upper righthand corner of the title page in preprint mode.
% Multiple \preprint commands are allowed.
% Use the 'preprintnumbers' class option to override journal defaults
% to display numbers if necessary
%\preprint{}

%Title of paper
\title{ Nesting and Degeneracy of Mie Resonances of Dielectric Cavities within Zero-Index Materials} 

\author{Xueke Duan$^1$}
\altaffiliation{These authors contributed equally to this work.}
\author{Haoxiang Chen$^1$}
\altaffiliation{These authors contributed equally to this work.}
\author{Yun Ma$^1$}
\altaffiliation{These authors contributed equally to this work.}
\author{ Zhiyuan Qian$^1$}
\author{ Qi Zhang$^1$}
\author{Yun Lai$^4$}
\author{Ruwen Peng$^4$}
\author{Qihuang Gong$^{1,2,3}$}
\author{Ying Gu$^{1,2,3}$}
 \email{ygu@pku.edu.cn}

\affiliation{$^1$State Key Laboratory for Mesoscopic Physics, Department of Physics, Peking University, Beijing 100871, China\\
$^2$Frontiers Science Center for Nano-optoelectronics $\&$  Collaborative Innovation Center of Quantum Matter $\&$ Beijing Academy of Quantum Information Sciences, Peking University, Beijing 100871, China\\
$^3$Collaborative Innovation Center of Extreme Optics, Shanxi University, Taiyuan, Shanxi 030006, China\\
$^4$ National Laboratory of Solid State Microstructures, School of Physics, and Collaborative Innovation Center of Advanced Microstructures, Nanjing University, Nanjing 210093, China\\
}

\date{\today}

\begin{abstract}

Resonances in optical cavities have been used to manipulate light propagation, enhance light-matter interaction, modulate quantum states, and so on.
However, in traditional  cavities, the permittivity contrast in and out the cavity is not so high.
Recently, zero-index materials (ZIMs) with unique properties and specific applications have attracted great interest.
By putting optical cavity into ZIMs,  the extreme circumstance with infinite permittivity contrast  can be obtained.  Here,  we theoretically study Mie resonances of dielectric cavities embedded in ZIMs with $\varepsilon \approx 0$, or $\mu \approx 0$, or $(\varepsilon,\mu) \approx 0$.
  Owing to ultrahigh contrast ratio of $\varepsilon$ or $\mu$ in and out the cavities, with fixed wavelength, 
a series of Mie resonances with the same angular mode number $l$  but with different cavity radii are obtained; more interestingly, its $2^l$-TM (TE) and $2^{l+1}$-TE (TM) modes have the same resonant solution for the cavity in $\varepsilon \approx 0$ ($\mu \approx 0$) material, and the resonance degeneracy also occurs between $2^l$-TM mode and $2^l$-TE mode for $(\varepsilon,\mu) \approx 0$ material.
We further use resonance degeneracy to modulate the Purcell effect of quantum emitter inside the cavity. 
The results of  resonance nesting and degeneracy will provide an additional view or freedom to enhance the performance of cavity behaviors.

%Nesting and Degeneracy of Resonances with ZIMs  .
%However, almost all of these studies only focus on the the zero index of bulk material itself and study the exotic behaviors of electromagnetic wave in and across ZIMs, and ignored the huge indexes contrast in and out the bulk ZIMs.
%Degenerate resonances own different linewidth, specifically the higher the $l$,  the narrower the linewidth.
%Zero-index materials (ZIMs) with $\varepsilon \approx 0$, or $\mu \approx 0$, or $(\varepsilon,\mu) \approx 0$ have attracted great interest due to their unique properties and specific applications.

\end{abstract}

% insert suggested keywords - APS authors don't need to do this
%\keywords{}

%\maketitle must follow title, authors, abstract, and keywords
\maketitle

% body of paper here - Use proper section commands
% References should be done using the \cite, \ref, and \label commands

%Introduction

Zero-index materials (ZIMs) \cite{Review2017,Science2013}, including  $\varepsilon$ near zero (ENZ), $\mu$ near zero (MNZ), and both $\varepsilon$ and  $\mu$ near zero (EMNZ) materials, have attracted great interest.
They have been experimentally realized in natural materials \cite{Optica2016,OME2011}, engineered dispersion waveguides \cite{WG1,WG2,WG3}, photonic crystals \cite{PC}, and metamaterials \cite{MM1,MM2,MM3,MM4}.
Owing to near zero $\varepsilon$  or $\mu$ \cite{Review2017,Science2013},  the electric field will decouple with the magnetic field in the ZIMs accompanied by constant phase distribution. 
With many attractive properties, like supercoupling \cite{Supercoupling1,Supercoupling2,Supercoupling3, WG3}, directional radiation phase pattern \cite{RPhase}, large optical nonlinearity  \cite{Non1,Non2},  random control of reflection and refraction \cite{TR1,TR2,TR3,TR4}, and resonance “pinning” effect \cite{Optica2016,Pinning2},  ZIMs have been used in coherent perfect absorption \cite{CPA1}, cloaking \cite{cloak}, waveguide connection \cite{Supercoupling1,Supercoupling2}, optical antennas \cite{Optica2016,Pinning2}, and so on. 
 However, these studies only focus on the zero index of bulk material itself rather than the huge index contrast in and out the bulk ZIMs.

%Recently, some researchers used the ZIM as environmental background and studied the geometry-invariant resonant cavities surrounded by ZIMs \cite{ENZcavity1,ENZcavity2,ENZcavity3,SE2}. 
%Resonant cavities have important applications in interaction between light and matter, sensing, optical  communication and so on. 
Optical cavities are ubiquitous, whose resonances can be used to manipulate light propagation, enhance light-matter interaction, modulate quantum states, and generate quantum sources.
With the contrast of  indexes ($\varepsilon, \mu$) in and out of the cavities,   optical responses such as surface plasmon resonance \cite{Metal1,Metal2,Gu2015} and dielectric resonance \cite{Die2,Die3,Die1,Die4,Die5} occur, characterized as strong local field enhancement. %For dielectric \cite{Die2,Die3,Die1,Die4,Die5} or metallic spherical cavities \cite{Metal1,Metal2} in the non-zero dielectric environment, series of resonances are perfectly figured out by Mie theory \cite{Mie,Debye1909,BOHREN1976}.
%The behaviors of high order resonances of dielectric cavities embedded in the ZIMs remain unknown. 
In  traditional  cavities, once the wavelength is fixed, resonance nesting with different size cavity and resonance degeneracy in nanoscale cavity, though which will provide additional degree of freedom to enhance the performance of photonic devices, have never been reported before. 
Despite with the highest contrast ratio of $\varepsilon$ or $\mu$ in and out the cavity embedded in the ZIMs,  only electric dipole resonance of  dielectric  cavity has been demonstrated  to modify the photon-emitter interaction \cite{ENZcavity1,ENZcavity2,ENZcavity3,SE2}. 

\begin{figure}[htbp]
\begin{center}
\includegraphics[width=0.3 \textwidth]{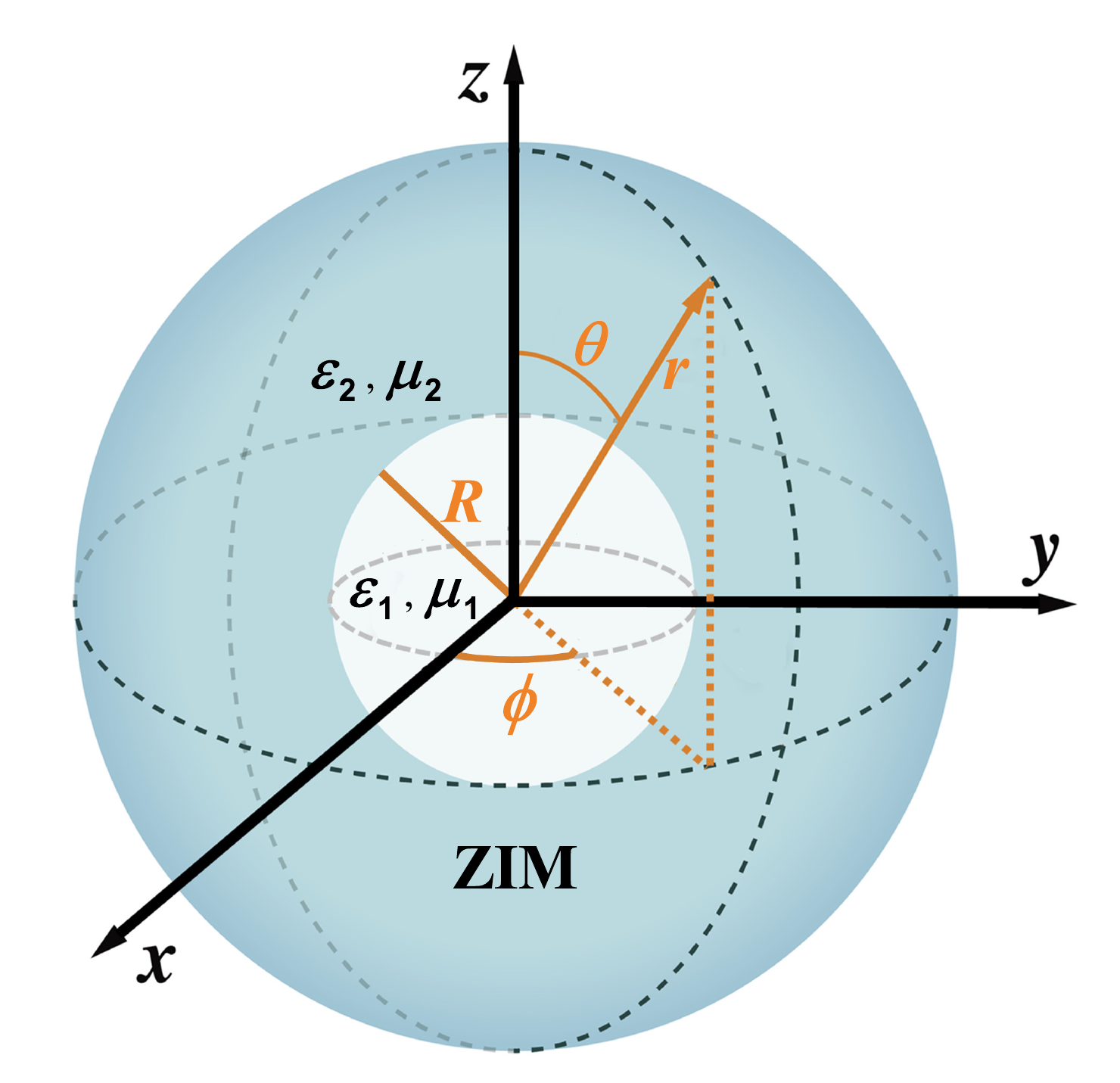}
\caption{\label{fig:system}The spherical cavity with zero-index background. The dielectric sphere (the white part) with radius of $R$ embedded in the infinite ZIM (the blue part). 
%The spherical coordinate system is established with the origin at the center point of the sphere. The dielectric constant of the sphere is $\varepsilon_1$ and the magnetic permeability is $\mu_1$. They are  $\varepsilon_2$ and  $\mu_2$ for the ZIM.
}
\end{center}
\end{figure}

Here, we analytically solve Mie resonances of dielectric spherical cavities embedded in ENZ, MNZ, and EMNZ background respectively [Fig.~\ref{fig:system}].  Unusually, for the same angular mode number $l$, a series of Mie resonances with different radii can be achieved at a fixed wavelength, so called resonance nesting.
More interestingly, the $2^l$-TM (TE) mode of the dielectric cavity has the same resonant frequency as that of its $2^{l+1}$-TE (TM) mode for the ENZ (MNZ) material; while  for EMNZ material, the resonance degeneracy occurs between its $2^l$-TM and $2^l$-TE modes.
The nesting and degeneracy of optical modes originate from the ultrahigh contrast ratio of $\varepsilon$ or $\mu$ in and out the cavities. Therefore, these phenomena also exist in nonspherical dielectric cavities surrounded by ZIMs.
We also find that  degenerate resonances own different linewidth, in other word, as the order $l$ becomes higher,  the linewidth becomes narrower. 
All above analytical results are confirmed by the numerical finite element method.
Owing to the resonance degeneracy of optical modes enabled by ZIMs, the interference or superposition between the modes is expected.
Then we used resonance degeneracy to modulate the photon-emitter interaction inside the cavity.
The resonance degeneracy and nesting enabled by zero-index materials  may have potential application in light manipulation, light-matter interaction, and photonic devices.

%%% Formulas part

The spherical cavity with ZIM background is shown in Fig.~\ref{fig:system}. The dielectric sphere (the white part) with the radius of  $R$ and dielectric constant $\varepsilon_1$ and magnetic permeability $\mu_1$  is embedded in the infinite ZIM (the blue part) with $\varepsilon_2$ and  $\mu_2$.  
In spherical coordinate system, different optical modes of dielectric spherical cavity are usually labeled as TM$_{lm}$/TE$_{lm}$ modes \cite{Mie,Debye1909,BOHREN1976,Die2,Die3,SM}, where TM  means transverse magnetic mode and TE means transverse electric mode; $l=1,2,3...$, the angular mode number, decides the polarity of the modes ($2^l$-modes), for example, $l=1$ means the dipole mode (2-mode) and $l=2$ means the quadrupole mode (4-mode); $m$ is the azimuthal mode number and satisfies $m \leq l$. As solving the Mie resonances of the cavity modes, we can take $m=0$ as an example because $m$ has no effect on the resonance conditions. On this premise, Mie resonances of cavity modes can be categorized into $2^l$-TM  and  $2^l$-TE Mie resonances. 
 
 %Commonly, the TE/TM modes with different $l$ are resonant with different conditions for both dielectric and metallic spherical cavities \cite{Die2,Die3,Die1,Die4,Die5,Metal1,Metal2}  embedded in non-zero materials. 
 
First considering the $2^l$-TM modes in the spherical cavity, because the magnetic field of the TM modes has no radial component, so the electromagnetic fields inside and outside the sphere can be written as \cite{SM}:
\begin{equation}
\begin{array}{l}
{\bf H}_{\mathrm {TM}}^l=\left\{\begin{array}{c}
{\bf M}_{l}^{(2)}+a {\bf M}_{l}^{(1)}, \quad r<R,\\
c {\bf M}_{l}^{(3)}, \quad r \geq R,\\
\end{array}\right.\\
{\bf E}_{\mathrm {TM}}^l=\left\{\begin{array}{c}
-\dfrac{k_{1}}{i \varepsilon_{1} \varepsilon_{0} \omega}\left({\bf N}_{l}^{(2)}+a {\bf N}_{l}^{(1)}\right), \quad r<R,\\
-\dfrac{k_{2}}{i \varepsilon_{2} \varepsilon_{0} \omega}\left(c {\bf N}_{l}^{(3)}\right), \quad r \geq R,
\end{array}\right.\\
\end{array}
\end{equation}
where $a$ and $c$ are coefficients to be determined,
$\bf M$ and $\bf N$ are two sets of Mie bases \cite{Mie,Debye1909,BOHREN1976} on which the electromagnetic field can be expanded. 
${\bf M}_{l}^{(j=1,2,3)}
=-\dfrac{\partial \mathrm{P}_{l}}{\partial \theta} z_{l}^{(j)}(x) \hat{\mathbf{e}}_{\phi}$, and
${\bf N}_{l}^{(j=1,2,3)}
=\dfrac{z_{l}^{(j)}(x)}{x} l(l+1) \mathrm{P}_{l} \hat{\mathbf{e}}_{r}+\dfrac{1}{x} \dfrac{\partial\left[x z_{l}^{(j)}(x)\right]}{\partial x}\dfrac{\partial \mathrm{P}_{l}}{\partial \theta} \hat{\mathbf{e}}_{\theta}$, in which $x=kr$ and $k$ is the wavenumber, labeled as $k_1$ in the sphere and $k_2$ out the sphere;
$z_{l}^{(j)}$ mean different kind of spherical harmonic functions respectively: spherical Bessel function $j_l$, spherical Neumann function $n_l$, and spherical Hankel function of the first kind $h_{l}^{(1)}$ which is a linear combination of $j_l$ and $n_l$, i.e. $h_{l}^{(1)}=j_l+i n_l$. For simplicity, we make $\eta_l(x) \equiv x j_l(x), \zeta_l(x) \equiv x n_l(x), \xi_l(x) \equiv x h_l^{(1)}(x)$. $P_{l}$ is the associated Legendre function. 
More details are shown in Ref. \cite{SM}. 

According to the continuous tangential electric field strength and magnetic field strength on the boundary ($r=R$), for the $2^l$-TM modes we get a linear equations with two coefficients $a$ and $c$ \cite{SM}:
 \begin{equation}
\left\{\begin{array}{l}
\tilde{\varepsilon}\left(\zeta_l^{\prime}(\rho)+a \eta_l^{\prime}(\rho)\right)=c \xi_l^{\prime}(s \rho), \\
\zeta_l(\rho)+a \eta_l(\rho)=c \dfrac{\xi_l(s \rho)}{s},
\end{array}\right.
\end{equation}
in which $\rho=k_1R$, $\tilde{\varepsilon}=\varepsilon_{2} / \varepsilon_{1}, \tilde{\mu}=\mu_{2} / \mu_{1}, s=k_{2} / k_{1}=\sqrt{\tilde{\varepsilon}\tilde{\mu}}$.  

%And the equations has the solution that:
%\begin{equation}
%a=\dfrac {\left|\begin{array}{cc}
%-\tilde{\varepsilon} \zeta_l^{\prime}(\rho) & -\xi_l^{\prime}(s \rho) \\
%-s\zeta_l(\rho) & -\xi_l(s \rho)
%\end{array}\right|} {\left|\begin{array}{cc}
%\tilde{\varepsilon} \eta_l^{\prime}(\rho) & -\xi_l^{\prime}(s \rho) \\
%s \eta_l(\rho) & -\xi_l(s \rho)
%\end{array}\right|}, 
%c=\dfrac {\left|\begin{array}{cc}
 %\tilde{\varepsilon} \eta_l^{\prime}(\rho)& -\tilde{\varepsilon} \zeta_l^{\prime}(\rho)\\
%s \eta_l(\rho) & -s\zeta_l(\rho) 
%\end{array}\right|} {\left|\begin{array}{cc}
%\tilde{\varepsilon} \eta_l^{\prime}(\rho) & -\xi_l^{\prime}(s \rho) \\
%s \eta_l(\rho) & -\xi_l(s \rho)
%\end{array}\right|}.
%\end{equation}
%

When the spherical cavity is resonant, $a$ and $c$ would go to extrema, which can be satisfied when the denominators of $a$ and $c$ are zero: 
\begin{equation}
\tilde{\varepsilon} \eta_l^{\prime}(\rho) \xi_l(s \rho)=s \eta_l(\rho)  \xi_l^{\prime}(s \rho).
\end{equation}
This is the limit situation for the resonance, whose premise of real solution can be naturally met by ZIMs with $s \approx 0$. In addition, with $s \approx 0$,  $\xi_l (s \rho) \approx a_l(s \rho)^{-l}$ and $\xi_{l}^{\prime}(s \rho) \approx (-l)a_l(s \rho)^{-(l+1)}$ \cite{SM},
substituting them into Eq. (3), we obtain:
\begin{equation}
\tilde{\varepsilon} \rho \eta_l^{\prime}(\rho)+l \eta_l(\rho)=0,
\end{equation}
\textit{which is the ideal Mie resonance condition for the $2^l$-TM modes of spherical cavity  embedded in ZIMs}.  Furthermore, for ENZ and EMNZ media, $\tilde{\varepsilon} \approx 0$, so Eq. (4) can be simplified to $\eta_l (\rho)=0$. Ideal resonance conditions only can be achieved when $s=0$ or $s$ is very near zero, but in fact, the small imaginary part of $\varepsilon_2$ or $\mu_2$ will make a little influence on the  $2^l$-TM Mie resonances \cite{SM}.

It is worth mentioning that in addition to ZIMs, $s\approx 0$ can also be satisfied by the situation that $\varepsilon_1 \gg \varepsilon_2$, i.e. the high index cavity embedded in low index material (like air). However, as discussed in Ref. \cite{SM}, the same resonant conditions as above  can be achieved only when $\varepsilon_1$ is very high (more than 900).  

For the $2^l$-TE modes in the spherical cavity, the electromagnetic fields inside and outside the sphere are written as  \cite{SM}:
\begin{equation}
\begin{array}{l}
{\bf E}_{\mathrm {TE}}^l=\left\{\begin{array}{c}
{\bf M}_{l}^{(2)}+b {\bf M}_{l}^{(1)}, \quad r<R,\\
d {\bf M}_{l}^{(3)}, \quad r \geq R,\\
\end{array}\right. \\
{\bf H}_{\mathrm {TE}}^l=\left\{\begin{array}{c}
\dfrac{k_{1}}{i \mu_{1} \mu_{0} \omega}\left({\bf N}_{l}^{(2)}+b {\bf N}_{l}^{(1)}\right), \quad r<R,\\
\dfrac{k_{2}}{i \mu_{2} \mu_{0} \omega}\left(d {\bf N}_{l}^{(3)}\right), \quad r \geq R,
\end{array}\right. \\
\end{array}
\end{equation}
where $b$ and $d$ are coefficients to be determined,

According to the continuous tangential electric field strength and magnetic field strength on the boundary ($r=R$), for the $2^l$-TE modes we can get  a linear equations with two unknown numbers $b$ and $d$ \cite{SM}:
 \begin{equation}
\left\{\begin{array}{l}
\tilde{\mu}\left(\zeta_l^{\prime}(\rho)+b \eta_l^{\prime}(\rho)\right)=d \xi_l^{\prime}(s \rho),\\
\zeta_l(\rho)+b \eta_l(\rho)=\dfrac{1}{s}d \xi_l(s \rho).
\end{array}\right.
\end{equation}
When in resonant, for $s\approx0$, the determinant of the coefficients of $b$ and $d$ should be zero, i.e.
\begin{equation}
\tilde{\mu} \eta_l^{\prime}(\rho) \xi_l(s \rho)=s \eta_l(\rho)  \xi_l^{\prime}(s \rho).
\end{equation}
Take further simplification of $\xi_l(s \rho)$, and we can get:
\begin{equation}
\tilde{\mu} \rho \eta_l^{\prime}(\rho)+l \eta_l(\rho)=0,
\end{equation}
\textit{which is the ideal Mie resonance condition for the $2^l$-TE modes of spherical cavity  embedded in ZIMs}. Specially, for MNZ and EMNZ media,  $\tilde{\mu} \approx 0$, so Eq. (8) can be simplified to $\eta_l(\rho)=0$. Similarly, the small imaginary part of $\varepsilon_2$ or $\mu_2$ will have effect on the $2^l$-TE Mie resonances but different with that on the $2^l$-TM Mie resonances \cite{SM}.

The Mie resonance conditions for $2^l$-TM and $2^l$-TE modes of dielectric spherical cavity placed in ENZ, MNZ, or EMNZ media are listed in Table.~\ref{tab:resonance1}. 
When the background varies from ENZ to EMNZ,  the resonance conditions of the $2^l$-TM modes have no change, but that of the $2^l$-TE modes are modulated and become the same as the $2^l$-TM modes when $\mu_2$ is also near zero. For the MNZ background, \textit{vice versa} \cite{SM}.

\begin{table}[htbp]
\renewcommand{\thetable}{1A}
\caption{\label{tab:resonance1}The Mie resonance conditions for $2^l$-TM and $2^l$-TE modes of dielectric spherical cavity embedded in ZIM.}

\begin{ruledtabular}
\begin{tabular}{cccc}
  & ENZ & MNZ& EMNZ \\
 %\hline
\colrule
$2^l$-TM mode & $\eta_l(\rho)=0$ & $\tilde{\varepsilon} \rho \eta_l^{\prime}(\rho)+l \eta_l(\rho)=0$ & $\eta_l(\rho)=0$ \\
$2^l$-TE mode &  $\tilde{\mu} \rho \eta_l^{\prime}(\rho)+l \eta_l(\rho)=0$ & $\eta_l(\rho)=0$  & $\eta_l(\rho)=0$ \\
\end{tabular}
\end{ruledtabular}
\renewcommand{\thetable}{1B}
\caption{\label{tab:resonance2}The Mie resonance conditions for 2-, 4- and  8- TM/TE modes of dielectric spherical cavity embedded in ZIM.}
\begin{ruledtabular}
\begin{tabular}{cccccccc}
 \footnote{A: $\sin \rho-\rho \cos \rho=0$; B: $\sin \rho=0$; C: $(3-\rho^2) \sin \rho-3\rho \cos \rho=0$; D: $(15-6\rho^2)\sin \rho-(15\rho-\rho^3)\cos \rho=0$} & 2-TM & 2-TE & 4-TM  &  4-TE  &  8-TM &  8-TE &... \\
\colrule
 %\hline
% ENZ\footnote{$(\mu_2=\mu_1)$} & A & B & C & A & D & C&...\\
  %MNZ\footnote{$(\varepsilon_2=\varepsilon_1)$} & B & A & A & C & C & D &... \\
   ENZ $(\mu_2=\mu_1)$ & A & B & C & A & D & C&...\\
  MNZ $(\varepsilon_2=\varepsilon_1)$ & B & A & A & C & C & D &... \\
EMNZ  & A & A & C & C & D & D &...\\
\end{tabular}
\end{ruledtabular}
\end{table}

 It can be seen that the resonance conditions are related to $\eta_l(\rho)$ and its derivative $\eta_l^{\prime}(\rho)$. Give the expression of $\eta_l(\rho)$ with $l=1,2,3$:
\begin{equation}
\begin{array}{c}
\eta_{1}(\rho)={\rho}^{-1}(\sin \rho-\rho \cos \rho), \\
\eta_{2}(\rho)={\rho^{-2}}\left[\left(3-\rho^{2}\right) \sin \rho-3 \rho \cos \rho \right], \\
\eta_{3}(\rho)={\rho^{-3}}\left[\left(15-6\rho^{2}\right) \sin \rho-\left(15 \rho-\rho^{3}\right) \cos \rho \right].
\end{array}
\end{equation}
Using the above formula and Table.~\ref{tab:resonance1}, we can get the resonance conditions for 2-, 4-, and 8- TE/TM modes, which are listed in Table.~\ref{tab:resonance2}. For conciseness, we use A to indicate $\sin \rho-\rho \cos \rho=0$, B to $\sin \rho=0$, C to $(3-\rho^2) \sin \rho-3\rho \cos \rho=0$ and D to $(15-6\rho^2)\sin \rho-(15\rho-\rho^3)\cos \rho=0$. More specially, if the refractive index $n=\sqrt{\varepsilon_1 \mu_1}$ of the sphere is 1, the resonant condition A can be replaced by $R/\lambda=0.7151,1.2295...$, B by $R/\lambda=0.5,1.0...$, C by $R/\lambda=0.9173,1.4475...$, and D by $R/\lambda=1.1122, 1.6579...$, where $\lambda$ is the wavelength in the vacuum.All above results are confirmed by the numerical finite element method \cite{SM}. While when the refractive index $n$ is not 1, the $nR/\lambda$ will be the above values when resonant.

%%%%%%%%%%%%%%%%%%%%%%%%%%%%%%%%%%%%%%%%%%%%%%%%%%%%%%

\begin{figure}
\begin{center}
\includegraphics[width=0.5 \textwidth]{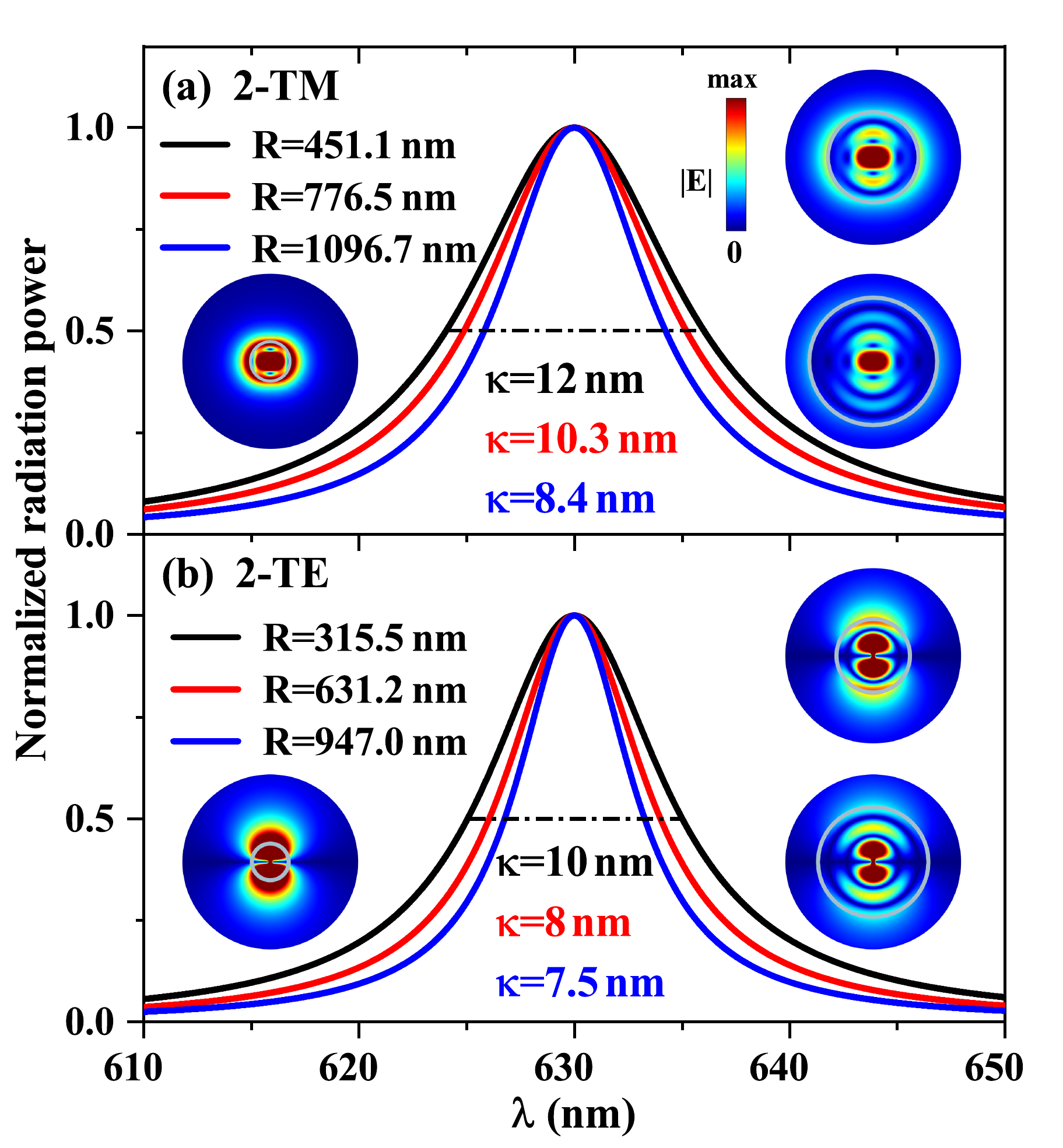}
\caption{\label{fig:nesting} Resonance nesting of (a) 2-TM and (b) 2-TE modes for the air sphere embedded in the ENZ medium when the resonant wavelength is fixed at 630 nm. 
%The peaks of the normalized radiation power spectra correspond to the resonance and the linewidth means the cavity loss $\kappa$.  2-TM resonance can be achieved at $R=451.1$ nm, 776.5 nm, 1096.7 nm, with $\kappa=12$ nm, 10.3 nm, 8.4 nm;  2-TE resonance can be gotten at $R=315.5$ nm, 631.2 nm, 947.0 nm, with $\kappa=10$ nm, 8 nm, 7.5 nm. Here,  $\varepsilon_2$ is set as $0.01i$, $\mu_2=1$, $\varepsilon_1$ and $\mu_1$ are both 1. 
The insets are corresponding electric field distributions with different $R$ (boundaries are shown as grey circles). Here,  $\varepsilon_2$ is set as $0.01i$, $\mu_2=1$.}
\end{center}
\end{figure}

%%% nesting part

% and the normalized form $\tilde{P}\propto \left | \dfrac{c}{s} \right |^2 \operatorname{Re} \left ( \dfrac{i}{s} \xi_l \xi_l^{'*} \right )$  \cite{SM}. The peak of the normalized radiation power corresponds to resonance and the linewidth means the cavity mode loss which is labeled as $\kappa$.  

\begin{figure}[htbp]
\begin{center}
\includegraphics[width=0.5 \textwidth]{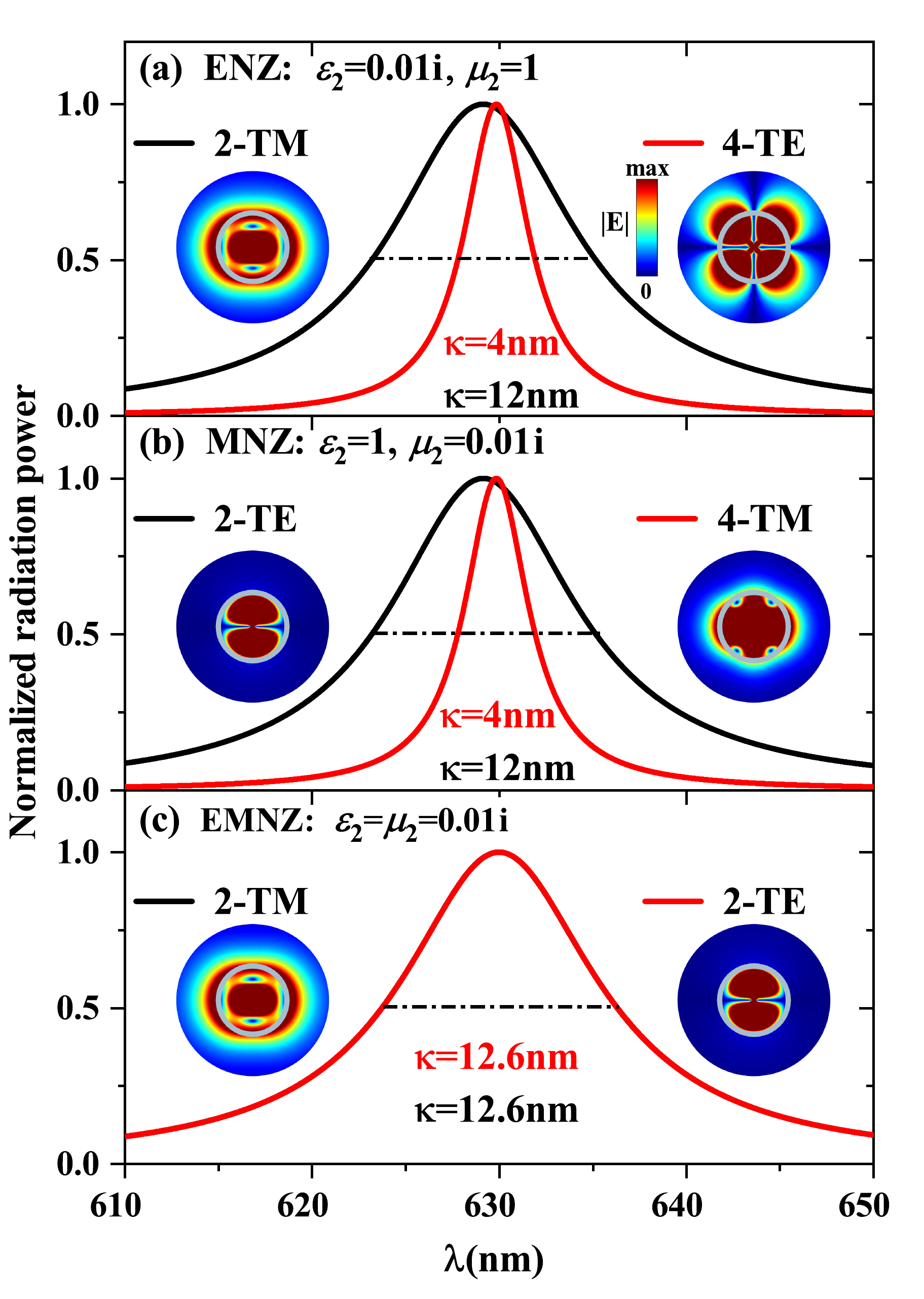}
\caption{\label{fig:degeneracy}
%The normalized radiation power spectra of the two degenerate modes of the air spherical cavity with $R=450.5$ nm embedded in different ZIM background: (a) 2-TM mode and 4-TE mode with $\varepsilon_2=0.01i$ and $\mu_2=1$, (b) 2-TE mode and 4-TM mode with $\varepsilon_2=1$ and $\mu_2=0.01i$, and (c) 2-TM mode and 2-TE mode with $\varepsilon_2=\mu_2=0.01i$. The cavity losses of the two degenerate modes are different in ENZ and MNZ cases, but are same in EMNZ case. The insets are the electric field distributions.
The normalized radiation power spectra of degenerate modes of air spherical cavity with $R=450.5$ nm in different ZIM: (a) 2-TM  and 4-TE modes in ENZ, (b) 2-TE and 4-TM modes in MNZ, and (c) 2-TM and 2-TE modes in EMNZ material. The insets are their electric field distributions.
}
\end{center}
\end{figure}

Different to plasmonic particles embedded in non-zero index media that usually have only one resonant $R/\lambda$ value for one mode \cite{Metal1,Metal2,Gu2015,DXK}, in the spherical cavity with ZIM background,  there are series of $R/\lambda$ values for each $2^l$-TM/TE Mie resonance. 
Namely, if the optical wavelength is fixed, the same Mie resonance can be achieved in spherical cavities with different radii $R$, which is called as ``resonance nesting''. 
As shown in Fig.~\ref{fig:nesting}, when the resonant wavelength is fixed at 630 nm ( take an example, also can be at other wavelengths \cite{SM}), the radiation power ($P =\oiint \frac{1}{2} \operatorname{Re}\left(\vec{E}^{*} \times \vec{H}\right) \cdot \mathrm{d} \vec{S}$ ) spectra of  2-TM resonance for ENZ case are analytically obtained at $R=451.1$ nm, 776.5 nm, 1096.7 nm..., and the spectra of 2-TE resonance  at $R=315.5$ nm, 631.2 nm, 947.0 nm.... 
 It can be seen from the insets of Fig.~\ref{fig:nesting} (a) (or (b)) that the electric field distributions of the three cavities are consistent in form, which just implies these cavities support the same kind resonance. While the values of cavity loss $\kappa$ are different, and the larger the cavity, the smaller the loss,  because of the increase of lossless energy storage space. 
 It is noted that, the resonant $R/\lambda$ values are little bigger than ideal values due to the imaginary part of $\varepsilon_2$, and approach ideal values with decreasing the imaginary part \cite{SM}.
Besides, the resonance nesting of 2-/4- modes for different ZIMs background is  shown in Ref. \cite{SM}.  
 
%%% degeneracy part

In addition to the nesting of the same polar mode, there is also the degeneracy between different polar modes. 
From Table.~\ref{tab:resonance2}, it can be seen that for the ENZ case when $\mu_1=\mu_2$, the $2^l$-TM and $2^{l+1}$-TE Mie resonances have the same resonance condition, i.e.  the same cavity can support both $2^l$-TM and $2^{l+1}$-TE modes at the same wavelength. 
Fig.~\ref{fig:degeneracy} (a) gives the normalized radiation power spectra of the 2-TM mode and 4-TE mode in the air cavity with $R=450.5$ nm embedded in ENZ background with $\varepsilon_2=0.01i$ and $\mu_2=1$.  
The little resonance shifts of the two modes originate from the effect of imaginary part of $\varepsilon_2$ \cite{SM}.
Furthermore, the values of $\kappa$ of the two degenerate modes are different, i.e., $\kappa=12$ nm for the 2-TM mode but $\kappa=4$ nm for the 4-TE mode. In a word, the same cavity support two modes with different loss: the higher the $l$,  the smaller the loss, due to less radiation.   
%Furthermore, the cavity loss of the degenerate modes are different, that is, the higher the $l$,  the smaller the loss, due to less radiation.   
%The insets are the electric filed distributions for 2-TM mode (left) and 4-TE mode (right). Because the electric field has the radial component for the TM mode, so the field has sudden change on the boundary.
The electric field distribution, for 2-TM mode, is discontinuous on the boundary due to the exist of radial component of $\bf E$ which suddenly changes with the high contrast ratio of $\varepsilon_2$ and $\varepsilon_1$;  but for 4-TE mode, the opposite is true \cite{SM}.
%{\color{red}As for electric field distribution, it is discontinuous on the boundary for 2-TM mode due to the exist of radial component of $\bf E$ which suddenly changes with the high contrast ration of $\varepsilon_2$ and $\varepsilon_1$;  but for 4-TE mode, opposite is true \cite{SM}.}

\begin{figure}[htbp]
\begin{center}
\includegraphics[width=0.5 \textwidth]{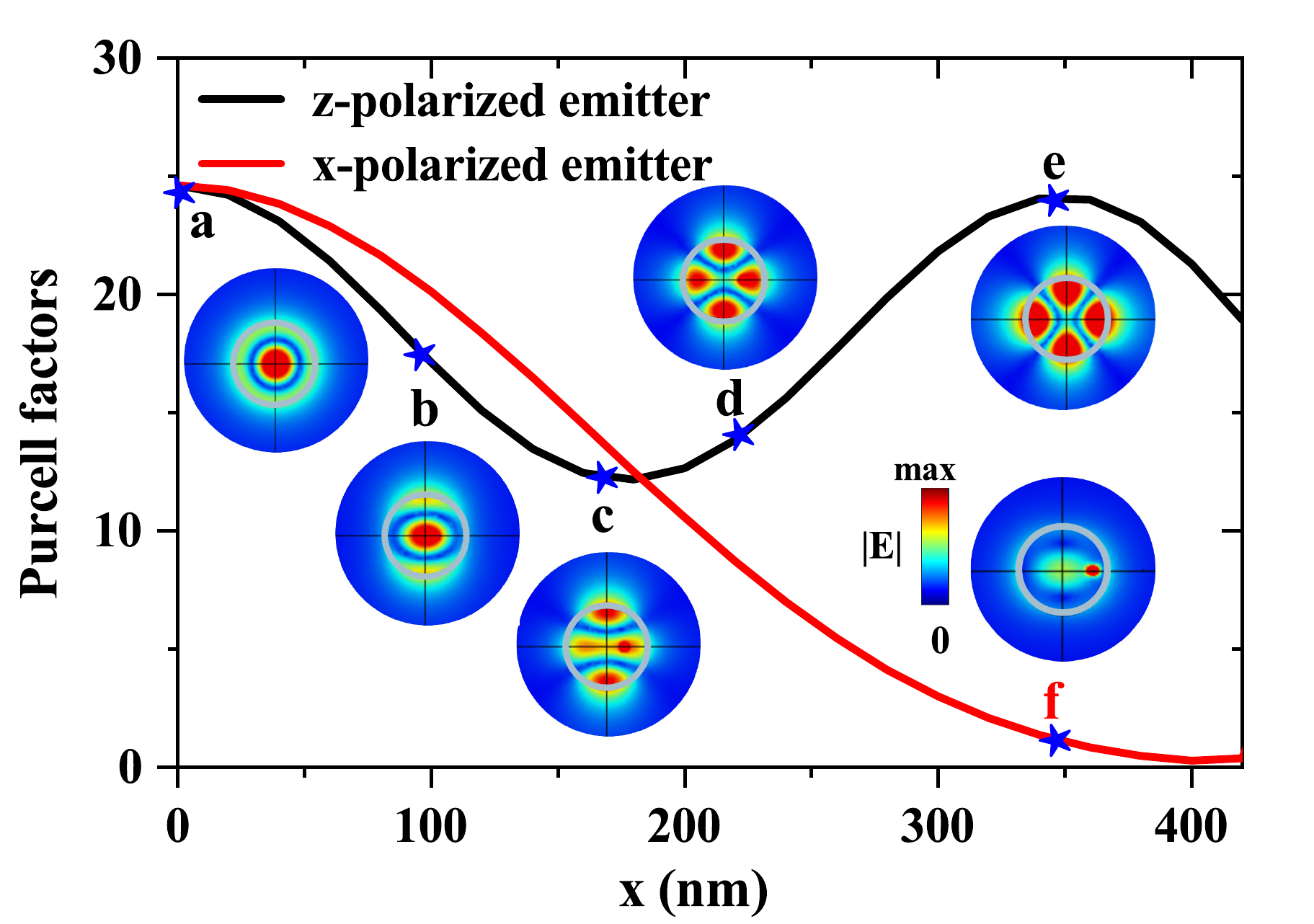}
\caption{\label{fig:PF} 
%The Purcell factors of the  $\parallel$-polarized or $\perp$-polarized dipole emitter as the emitter moving along the x-axis in the air cavity with ENZ background.  Here, $R=451.1$ nm, $\varepsilon_2=0.01i$ and $\mu_2=1$.  From the electric field distributions shown in insets, it can be seen that for the $\perp$-polarized emitter, the 2-TM mode (point a) or 4-TE (d and e) mode or both of them (b and c) can be inspired when the emitter at different positions, and the Purcell factor can keep above 10 and experiences two maximum values; while for the $\parallel$-polarized emitter, only the 2-TM mode can be excited with the moving of the emitter (a and f), and the Purcell factor decrease to zero near the boundary.
The Purcell factors of the  $z$-polarized or $x$-polarized dipole emitter moving along the $x$-axis in the air cavity with ENZ background.  
The electric field distributions of points a-f are shown in insets.
%Here, the parameters are the same as those in Fig. 2.  
Here, $R=451.1$ nm, $\varepsilon_2=0.01i$, and $\mu_2=1$.
}
\end{center}
\end{figure}

The resonant degeneracy also happens between the $2^l$-TE and $2^{l+1}$-TM Mie resonances for the MNZ case when $\varepsilon_1=\varepsilon_2$  [Table.~\ref{tab:resonance2}]. 
As shown in Fig.~\ref{fig:degeneracy} (b), the normalized radiation power spectra of the $2^l$-TE (TM) mode for the MNZ case are same with that of the $2^l$-TM (TE) mode for the ENZ case, because of the symmetry of electromagnetic field expressions.
In the same way, the little difference between the resonant wavelength of the 2-TE and 4-TM modes caused by the influence of imaginary part of $\mu_2$ (here $\mu_2=0.01i$ and $\varepsilon_2=1$).
The electric field distribution, no matter for 2-TE or 4-TM mode, is continuous on the boundary because $\varepsilon_2=\varepsilon_1$; and specially for the 2-TE mode, the electric field is almost zero out the sphere.
%because the $\bf E$ is near zero on the boundary under the requirement of resonance and continuous in and out the sphere; but for the 4-TM mode, the electric field is just continuous on the boundary with no .
%because the electric field near the boundary in the cavity is reduced to zero and it is continuous with only tangential components; but not for the 4-TM mode and the electric field is continuous on the boundary.}
The magnetic field distribution of the $2^l$-TM mode of ENZ case is same with the electric distribution of the $2^l$-TE mode of the MNZ case, and \textit{vice versa}. 

For the EMNZ case, the resonant degeneracy occurs between the $2^l$-TM and $2^l$-TE modes.
% because the  $\varepsilon_2$ and $\mu_2$ are both near zero.
It can be seen from the Fig.~\ref{fig:degeneracy} (c) that the normalized radiation power spectra for 2-TM and 2-TE modes overlaps together with the same cavity loss $\kappa=12.6$ nm. The electric field distribution of the 2-TM mode has the same form with that in the ENZ case and the 2-TE mode is similar with that in the MNZ case. 
Although resonance conditions of $2^l$-TM mode for ENZ case, $2^l$-TE mode for MNZ case and $2^l$-TM mode for EMNZ case have the same form, but they can not be regarded as degenerate because of the different electromagnetic backgrounds.

%%% discussion part
Next, we used resonance degeneracy to study the photon-emitter interaction. Here, we take the air cavity  with radius $R=451.1$ nm embedded in ENZ medium as an example. Choosing the surface where the electric field strength out the cavity is reduced to $1/e$ of that on the boundary,  we estimate that the mode volume is about $11R^3$,  and then the coupling strength $g$  is no more than 1meV when the transition dipole moment is 0.5 enm \cite{SM,Gu2017}, much lower than the cavity loss $\kappa$ which is about  37 meV for 2-TM mode and 12.5 meV for 4-TE mode \cite{SM}. So the interaction between the photon and the dipole emitter is at the weak coupling region in this system, and then we study the Purcell effect \cite{PF} of the emitter. 

We build spherical module in the COMSOL Multiphysics software to calculate Purcell factors by the the ratio of the radiated power of the emitter in the cavity and that of the emitter in the vacuum \cite{SM}. 
%For ENZ case, the parameters are the same as those in Fig. 2. 
When the dipole emitter is at the center of the cavity, only the 2-TM mode can be excited, but the 4-TE mode can be inspired when the emitter moves  to the edge  [Fig.~\ref{fig:PF}].
For the $z$-polarized emitter,  the 2-TM mode (point a ) or 4-TE mode (point e and d) or both of them (points b and c) can be excited and the Purcell factor can keep above 10.  It experiences two maximum values due to the interference or superposition between 2-TM mode and 4-TE modes.
While for the $x$-polarized emitter, no matter where the emitter is, only the 2-TM mode is excited  and the Purcell factor decrease to zero near the boundary. 
In addition, the Purcell factors will increase greatly if the imaginary part of the background decreases \cite{SM}.

%%% summary 
%We have theoretically demonstrated resonance nesting, i.e., for the same angular number $l$ and resonant wavelength, there is a series of $2^l$-TE or $2^l$-TM resonances {\color{red}occurring at} different cavity radii. 
%{\color{red}More significantly, we have also revealed the phenomenon of }resonance degeneracy between the $2^l$ -TM (TE) mode and the $2^{l+1}$ -TE (TM) mode {\color{red}of the spherical cavity} for the  ENZ (MNZ) background, as well as between $2^l$-TM mode and $2^l$-TE mode for the EMNZ background. 
%{\color{red}All above results have been confirmed by the numerical finite element method  \cite{SM}, and the resonance nesting can be achieved at other }
%Furthermore, we have used the degenerate resonances to modulate the photon-emitter interaction inside the cavity. 
In summary,  we have analytically solved the Mie resonances of dielectric spherical cavities embedded in the ENZ, MNZ, and EMNZ materials. 
We have theoretically revealed the phenomena of resonance nesting and resonance degeneracy existing in zero-index materials.
The  nesting and degeneracy originate from the high contrast ratio of $\varepsilon$ or $\mu$ in and out the cavities, thus if the cavities with large $\varepsilon$ or $\mu$ embedded in the low index materials, the same phenomena will occur \cite{SM}. 
Owing to possessing  the same physical principle, resonance nesting will also exist in other geometrical cavities embedded in ZIMs and  resonance degeneracy will occur in spherical symmetrical cavities because of the common spherical harmonics.
In contrast to previous mode degeneracy generally occurring between $+l$ and $-l$,  the mode degeneracy here with different angular mode number $l$ will provide an additional way to realize quantum entanglement and quantum operation.
The resonance degeneracy enabled by ZIMs may have potential application in light manipulation, light-matter interaction, and photonic devices.

\acknowledgments
This work is supported by the National Key R$\&$D Program of China under Grant No. 2018YFB1107200, by the National Natural Science Foundation of China under Grants Nos. 11525414,  11974032,  11734001, and  11974176,  and by the Key R$\&$D Program of Guangdong Province under Grant No. 2018B030329001.

\bibliography{ZIM}

\end{document}